\documentclass[aps,prd,twocolumn,showpacs,amsmath,nofootinbib]{revtex4-1}

\usepackage{color}
\newcommand{\beqa}{\begin{eqnarray}}
\newcommand{\eeqa}{\end{eqnarray}}
\usepackage{float}
\usepackage{epsfig}
\usepackage{graphicx}
\usepackage{latexsym}

\usepackage{amsmath}
\usepackage{mathrsfs}

\usepackage{dcolumn}
\usepackage{bm}%

\begin{document}

\title{Constraints on accreting primordial black holes with the global 21-cm signal}

\author{Yupeng Yang}

\affiliation{School of Physics and Physical Engineering, Qufu Normal University, Qufu, Shandong, 273165, China}

\begin{abstract}
We investigate the effect of accreting primordial black holes (PBHs) on the thermal history of the intergalactic medium (IGM), including the accretion of baryonic matter and dark matter particles. The variations of the thermal history of the IGM caused by accreting PBHs will  result in the changes of the global 21-cm signal in the cosmic dawn. Based on the detection of the global 21-cm signal by EDGES, by requiring the differential brightness temperature, e.g., $\delta T_{21} \lesssim -100~(-50)~\rm mK$, we obtain the upper limits on the abundance of PBHs for the mass range $10\lesssim M_{\rm PBH} \lesssim 10^{4}~M_{\odot}$. 
Compared with previous works, the limits are stronger for the mass range $10\lesssim M_{\rm PBH}\lesssim 50~M_{\odot}$. 	

\end{abstract}

\maketitle

\section{introduction} 
The Experiment to Detect the Global Epoch of Reionization Signature (EDGES) has recently reported the detection of the 
global 21-cm signal in the cosmic dawn, finding an abnormal absorption signal with an amplitude of $T_{21}\sim 500\rm~mK$ centered at redshift $z\sim 17$~\cite{edges-nature}. The maximum amplitude of the detected global 21-cm signal is about twice larger than expected in the standard scenario~\cite{Xu:2021zkf,Furlanetto:2006jb,Cohen:2016jbh}. Some schemes have been proposed to explain the detected abnormal signal 
(see, e.g., Refs.~\cite{Feng:2018rje,Barkana:2018cct}). On the other hand, the abnormal signal has been used to investigate the  properties of dark matter (DM)~\cite{Hektor:2018lec,Kovetz:2018zan,Barkana:2018cct,prd-edges,Hektor:2018qqw,yinzhema,PhysRevD.103.123002,Hiroshima:2021bxn,Halder:2021rbq,Cheung:2018vww,Fraser:2018acy,Berlin:2018sjs,Fialkov:2018xre}.   

Although DM has been confirmed by many different astronomical observations, its nature is still unknown. Many DM models have been proposed 
and the most researched model is the weakly interacting massive particles (WIMPs)~\cite{Bertone:2004pz,Jungman:1995df}. Up until now, no detection of WIMPs has led to renewed interest in other DM models, such as primordial black holes (PBHs)~\cite{pbhs_review,Cai:2020fnq,Villanueva-Domingo:2021spv,Green:2020jor}. In particular, it has been argued that gravitational waves detected recently by LIGO and Virgo could be partially caused partly by the mergers of PBHs~\cite{Bird:2016dcv,Clesse:2020ghq,Deng:2021ezy,Franciolini:2021tla}. Moreover, the abundance of PBHs can be constrained from studies on the detected gravitational waves~\cite{Kohri:2020qqd,Wang:2016ana,Hutsi:2020sol}.

PBHs can form in the early universe via the collapse of large density perturbations~\cite{carr,pbhs_review}. When PBHs form,  baryonic matter is accreted onto them with emission of high energy photons during the process of accretion~\cite{Ricotti:2007jk,Ricotti:2007au,Poulin:2017bwe,Ali-Haimoud:2016mbv}. The interactions between the high energy photons emitted from accreting PBHs and the particles 
existing in the Universe will lead to the changes of the thermal history of the intergalactic medium (IGM), which will be reflected in the global 21-cm signal~\cite{mnras,prd-edges,yinzhema,Ricotti:2007au,Cang:2020aoo,Poulin:2017bwe,Ali-Haimoud:2016mbv,Tashiro:2012qe,Mittal:2021egv,Natwariya:2021xki,Cang:2021owu,Mena:2019nhm,Villanueva-Domingo:2021cgh}.

In addition to the accretion of baryonic matter, DM particles can also be accreted if PBHs do not make up all DM. Both theoretical 
research and simulation have shown that a DM halo surrounding a PBH with a density profile $\rho_{\rm DM}(r)\sim r^{-9/4}$ should be formed due to accretion~\cite{josan,0908.0735,Cai:2020fnq,Adamek:2019gns,Eroshenko:2016yve}. Compared to the case with no DM halo, the accretion rate of PBHs with a DM halo is enhanced~\cite{Ricotti:2007au,Ricotti:2007jk}. The influence of accreting PBHs without DM accretion on the evolution of the IGM have been investigated, and constraints on the abundance of PBHs are obtained using the cosmic microwave background (CMB) data~\cite{Ali-Haimoud:2016mbv,Poulin:2017bwe,Ricotti:2007au,Chen:2016pud} and the global 21-cm signal detected by EDGES~\cite{Hektor:2018qqw}. 
The authors of Ref.~\cite{PhysRevResearch.2.023204} investigated the constraints on the abundance of PBHs using the Planck data including 
the accretion of DM particles. 
In this paper, by taking into account the accretion of baryonic matter and DM particles onto PBHs, we investigate the effects of accreting PBHs on the thermal history of the IGM and the global 21-cm signal in the cosmic dawn. By requiring the differential brightness temperature,  e.g., $\delta T_{\rm 21}\lesssim\rm -100~(-50)~mK$, we obtain the upper limits on the abundance of PBHs for the mass range $10\lesssim M_{\rm PBH} \lesssim 10^{4} M_{\odot}$.

This paper is organized as follows. In Sec. II we discuss the basic properties of accreting PBHs. In Sec. III, the influence of accreting PBHs on the thermal history of the IGM and the global 21-cm signal in the cosmic dawn are investigated, and then the upper limits on the abundance of PBHs are obtained. The conclusions are given in Sec. IV.


\section{The basic properties of accreting PBHs}
In this section we review the basic properties of accreting PBHs and one can refer to, e.g., Refs.~\cite{Ricotti:2007au,Ricotti:2007jk,Ali-Haimoud:2016mbv} for more details. A PBH with mass $M_{\rm PBH}$ can accrete surrounding baryonic matter at the Bondi-Hoyle rate $\dot{M}_{\rm HB}$ as follows~\cite{Poulin:2017bwe,Ricotti:2007au}: 

\beqa
\dot{M}_{\rm HB} = 4\pi \lambda \rho_{\infty}(GM_{\rm PBH})^2 v^{-3}_{\rm eff},
\eeqa 
where $G$ is the gravitational constant. The mass density of cosmic gas far away from PBHs is $\rho_{\infty}=n_{\infty}m_{p}$, where $m_p$ is the proton mass and $n_{\infty}$ is the mean cosmic gas density. $\lambda$ is the accretion parameter taking into account the effects of gas viscosity, Hubble expansion, and Compton scattering between the CMB and cosmic gas~\cite{Ricotti:2007au,Ricotti:2007jk,Ali-Haimoud:2016mbv}. The effective velocity $v_{\rm eff}$ involves the gas sound speed ($c_{s,\infty}$) and the relative velocity ($v_{R}$) between PBHs and baryons. The gas sound speed which is far away from PBHs, $c_{s,\infty}$, can be written as follows~\cite{Poulin:2017bwe}: 

\beqa
c_{s,\infty} =\sqrt{\frac{\gamma (1+x_{e})T}{m_{p}}},
\eeqa
where $\gamma=5/3$ and $x_e$, $T$ and $m_p$ are the ionization fraction, the temperature of baryons and proton mass respectively. In the linear regime, the square root of the variance of relative velocity $v_{R}$ is given by~\cite{Poulin:2017bwe,Ricotti:2007au}

\beqa
\left<v_{R}^{2}\right>^{1/2} \approx {\rm min}\left[1,\frac{1+z}{1000}\right]\times 30~\rm km~s^{-1}.
\eeqa

The energy released from accreting PBHs and injected into the IGM should be obtained by averaging the luminosity of accreting PBHs over the Gaussian distribution of relative velocities~\cite{Ali-Haimoud:2016mbv,Poulin:2017bwe,Ricotti:2007au}. It has been proposed that the effective velocity can be defined as $v_{\rm eff}\equiv \left<\left(c^{2}_{s,\infty}+v_{R}^{2}\right)^{-3}\right>^{-1/6}$ with an approximate form as follows~\cite{Ali-Haimoud:2016mbv,Poulin:2017bwe,Hasinger:2020ptw}:

\begin{equation}
{v_{\rm eff}}\approx
\begin{cases}
\sqrt{c_{s,\infty}\left<v_{R}^{2}\right>^{1/2}}  ~~{\rm for}~c_{s,\infty}\ll \left<v_{R}^{2}\right>^{1/2} \\
c_{s,\infty} ~~~~~~~~~~~~~~~{\rm for}~c_{s,\infty}\gg \left<v_{R}^{2}\right>^{1/2}
\end{cases}
\end{equation}

The accretion luminosity of a PBH, $L_{\rm acc, PBH}$, is proportional to the Bondi-Hoyle rate $\dot M_{\rm HB}$~\cite{Poulin:2017bwe}:

\beqa
L_{\rm acc, PBH} = \epsilon \dot{M}_{\rm HB}c^2,
\eeqa
where $\epsilon$ is the radiative efficiency. It depends on the details of accretion and a typical value 
$\epsilon=0.01\dot m$ is usually used for spherical accretion~\cite{Ricotti:2007au}. $\dot{m}$ is the dimensionless Bondi-Hoyle accretion rate defined as $\dot{m} = \dot{M}_{\rm HB}c^{2}/L_{\rm Edd}$, where $L_{\rm Edd}=1.26\times 10^{38}\left(M/M_{\odot}\right)~\rm erg~s^{-1}$ is the Eddington luminosity. The authors of Ref.~\cite{Ali-Haimoud:2016mbv} reanalyzed the accretion process of PBHs and found $\epsilon=10^{-5} (10^{-3})\dot{m}$ for collisional ionization (photoionization). In this work, in order to investigate the conservative upper limits on the abundance of PBHs, we use 
$\epsilon=10^{-5}\dot{m}$ for our calculations and a larger value of $\epsilon$ will strengthen the final limits. 
Using the formulas above, the dimensionless Bondi-Hoyle accretion rate of a PBH without DM particle accretion can be written as follows~\cite{Ricotti:2007au}

\beqa
\dot{m}= 0.4\lambda \left(\frac{1+z}{1000}\right)^{3}\left(\frac{M_{\rm PBH}}{M_{\odot}}\right)\left(\frac{v_{\rm eff}}{\rm km~s^{-1}}\right)^{-3}.
\label{eq:m_nohalo}
\eeqa

Following Ref.~\cite{Ricotti:2007au}, we use the fitted formula of $\lambda$ as follows

\beqa
\lambda = {\rm exp}\left(\frac{4.5}{3+\beta^{0.75}}\right) x_{\rm cr}^2,
\label{eq:lambda}
\eeqa
where $x_{\rm cr}\equiv r_{\rm cr}/r_{B}$ is the dimensionless sonic radius and can be written in the form: 

\beqa
x_{\rm cr} = \frac{-1+\sqrt{1+\beta}}{\beta},
\eeqa
where $\beta$ is the dimensionless viscosity~\cite{Ali-Haimoud:2016mbv,Ricotti:2007jk,Ricotti:2007au}. 

If PBHs do not constitute all DM, in addition to baryonic matter, DM particles can also be accreted onto PBHs. Both simulation and theoretical research have shown that a DM halo surrounding a PBH can be formed due to accretion, and the density profile of a DM halo is $\rho_{\rm DM}(r)\sim r^{-\alpha}$ with $\alpha = 2.25$~\cite{josan,0908.0735,Cai:2020fnq,Adamek:2019gns,Eroshenko:2016yve}. For a PBH 
with $M_{\rm PBH}$, the mass of a DM halo surrounding PBH is changed with redshift as~\cite{0908.0735,Ricotti:2007jk}

\beqa
M_{\rm halo}=3M_{\rm PBH}\left(\frac{1+z}{1000}\right)^{-1}.
\eeqa

Taking the effect of the DM halo into account, the dimensionless Bondi-Hoyle accretion rate is written as follows~\cite{Ricotti:2007au,DeLuca:2020fpg}:

\beqa
\dot{m}=3\lambda\left(\frac{1+z}{1000}\right)\left(\frac{M_{\rm PBH}}{M_{\odot}}\right)\left(\frac{v_{\rm eff}}{\rm km~s^{-1}}\right)^{-3}.
\eeqa

Moreover, the relevant parameters $\beta$, $\lambda$, and $r_{\rm cr}$ should be rescaled as follows~\cite{{Ricotti:2007au,Ricotti:2007jk,DeLuca:2020fpg}} 

\beqa
\beta^{'} \equiv \kappa^{\frac{p}{1-p}}\beta,~\lambda^{'} \equiv \Upsilon^{\frac{p}{1-p}}\lambda(\beta^{'}),~r_{\rm cr}^{'}\equiv \left(\frac{\kappa}{2}\right)^{\frac{p}{1-p}}r_{\rm cr}, 
\label{eq:p_halo}
\eeqa
where $p=3-\alpha$ and $r_{\rm cr} = x_{\rm cr}r_{B}$, $r_{B} \equiv GM/v_{\rm eff}^{2}$ is the Bondi-Hoyle radius. The parameters $\kappa$ and $\Upsilon$ are given by~\cite{Ricotti:2007au,DeLuca:2020fpg}

\beqa
\kappa \equiv \frac{r_{\rm B}}{r_{\rm halo}} =  0.22\left(\frac{1+z}{1000}\right)\left(\frac{M_{\rm halo}}{M_{\odot}}\right)^{2/3}\left(\frac{v_{\rm eff}}{\rm km~s^{-1}}\right)^{-2}
\label{eq:kappa}
\eeqa

\beqa
\Upsilon = \left(1+10\beta^{'}\right)^{1/10}{\rm exp}~\left(2-\kappa\right)\left(\frac{\kappa}{2}\right)^{2}
\label{eq:upsilon}
\eeqa

For $\kappa \geq 2$, the DM halo surrounding a PBH is treated as a point mass, while the relevant parameters should be modified with Eqs.~(\ref{eq:p_halo}), (\ref{eq:kappa}) and (\ref{eq:upsilon}) for $\kappa < 2$. In this paper, we assume that the bolometric luminosity is lower than the Eddington limits. For the feedback effect, we follow the discussions given in Ref.~\cite{Ricotti:2007au}.


\section{The evolution of the IGM and the global 21-cm signal in the cosmic dawn including accreting PBHs}

\subsection{The evolution of the IGM including accreting PBHs}

The interactions between the high energy photons emitted from accreting PBHs and the particles existing in the Universe, lead to the changes of the degree of ionization and the thermal history of the IGM through heating, ionization, and excitation~\cite{lz_decay,xlc_decay,yinzhema,mnras,DM_2015,prd-edges,Belotsky:2014twa}. The changes of the degree of ionization ($x_e$) and the temperature of the IGM ($T_{k}$) with redshift are governed by the following equations~\cite{mnras,DM_2015,xlc_decay,lz_decay}: 

\beqa
(1+z)\frac{dx_{e}}{dz}=\frac{1}{H(z)}\left[R_{s}(z)-I_{s}(z)-I_{\rm PBH}(z)\right],
\label{eq:xe}
\eeqa

\beqa
(1+z)\frac{dT_{k}}{dz}=\frac{8\sigma_{T}a_{R}T^{4}_{\rm CMB}}{3m_{e}cH(z)}\frac{x_{e}(T_{k}-T_{\rm CMB})}{1+f_{\rm He}+x_{e}}
\\ \nonumber
-\frac{2}{3k_{B}H(z)}\frac{K_{\rm PBH}}{1+f_{\rm He}+x_{e}}+T_{k}, 
\label{eq:tk}
\eeqa
where $R_{s}(z)$ and $I_{s}(z)$ are the recombination and ionization rate from the standard sources, respectively. 
The ionization rate ($I_{\rm PBH}$) and heating rate ($K_{\rm PBH}$) caused by accreting PBHs can be written as follows~\cite{lz_decay,xlc_decay,mnras,yinzhema,DM_2015}: 

\beqa
I_{\rm PBH} = f(z)\frac{1}{n_b}\frac{1}{E_0}\frac{{\rm d}E}{{\rm d}V{\rm d}t}\bigg|_{\rm PBH} 
\label{eq:I}
\eeqa
\beqa
K_{\rm PBH} = f(z)\frac{1}{n_b}\frac{{\rm d}E}{{\rm d}V{\rm d}t}\bigg|_{\rm PBH} 
\label{eq:K}
\eeqa
where $n_b$ is the number density of baryon and $E_0= 13.6~\rm eV$. $f(z)$ is the energy fraction injected into the IGM for ionization, heating and exciting, respectively. It depends on redshift and has been analyzed in detail (see, e.g., Refs.~\cite{energy_function,Slatyer:2015kla,Poulin:2017bwe}). The main factor determining $f(z)$ is the spectrum of the radiation from accreting PBHs. For the spherical accretion considered here, the radiation is mainly from the bremsstrahlung 
emission~\cite{spectrum_1,spectrum_2,Ali-Haimoud:2016mbv,Poulin:2017bwe}, and the spectrum can be written as~\cite{Poulin:2017bwe}:

\beqa
L_{\omega}\propto \omega^{-a}{\rm exp}(-\omega/T_{s})
\eeqa 
where $T_{s}\sim 0.2~\rm MeV$ and $|a|\lesssim 0.5$. For our purposes, we use the public code ExoCLASS~\cite{exoclass}, 
which is a branch of the public code CLASS~\cite{class} and uses the above spectrum, to calculate $f(z)$ numerically. 

The energy injection rate per unit volume from accreting PBHs can be written as follows~\cite{Poulin:2017bwe}:

\beqa
\frac{{\rm d}E}{{\rm d}V{\rm d}t}\bigg|_{\rm PBH} =L_{\rm acc,PBH}f_{\rm pbh}\frac{\rho_{\rm DM}}{M_{\rm PBH}},
\eeqa 
where $f_{\rm pbh}=\rho_{\rm PBH}/\rho_{\rm DM}$. One should note that a monochromatic mass distribution for PBHs has been adopted for our calculations.

The changes of the temperature of the IGM, $T_k$, with redshift can be obtained by solving the differential equations, ~(\ref{eq:xe})-(\ref{eq:K}). Specifically, we have modified the public code RECFAST in CAMB\footnote{https://camb.info/}, including the effects from accreting PBHs, to solve the differential equations numerically~\cite{mnras,DM_2015,xlc_decay,lz_decay,prd-2020,yinzhema}. 
It should be noticed that the fudge factors are used in RECFAST module to calibrate for the standard cosmology, and more accurate calculations can be found from CosmoRec~\cite{Chluba:2010ca} and HyRec~\cite{Ali-Haimoud:2010hou} modules. For our purposes, following Ref.~\cite{Poulin:2017bwe}, it is enough to use RECFAST for the calculations.

\begin{figure}
\centering
\includegraphics[width=0.5\textwidth]{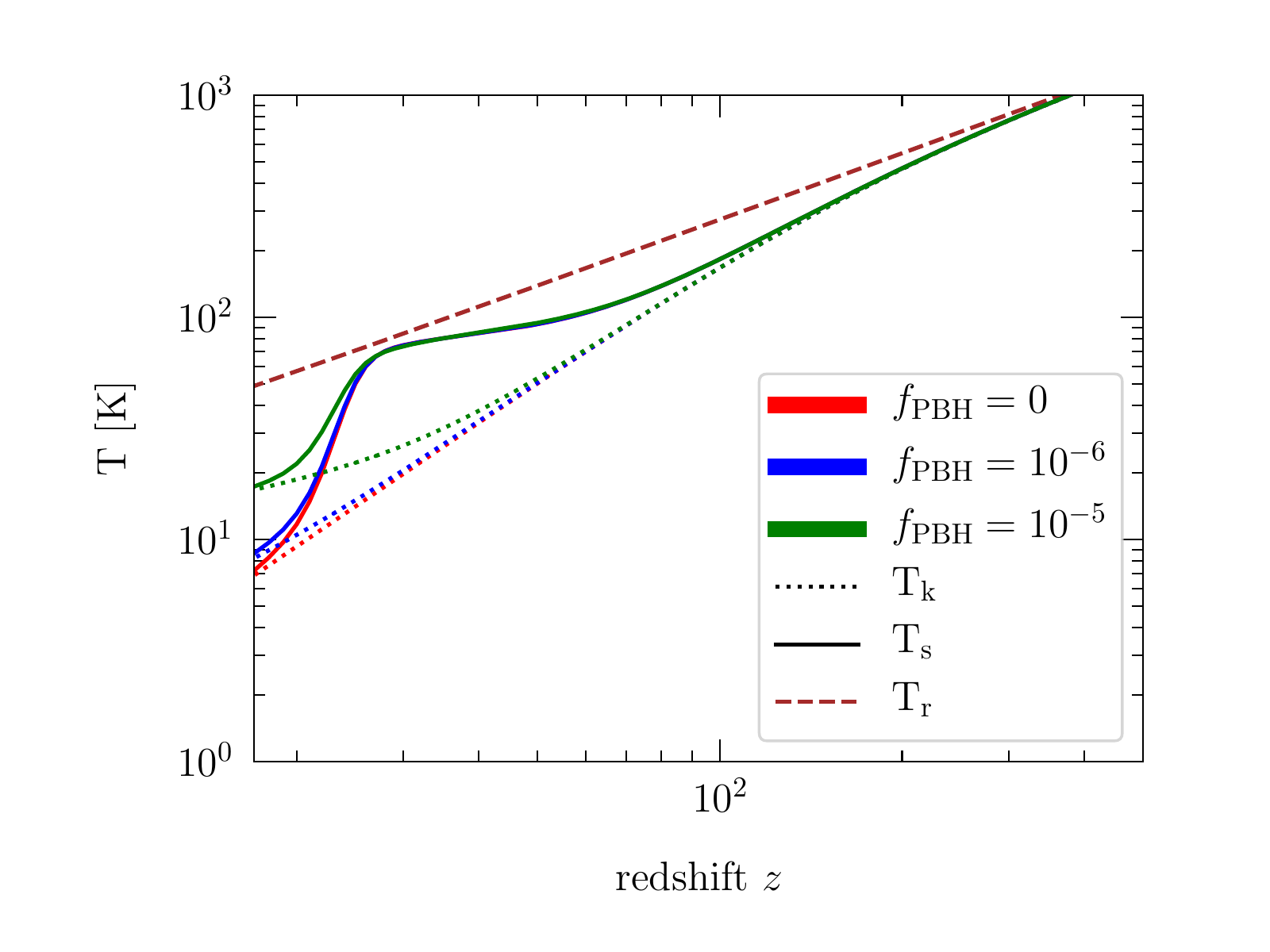}
\caption{The gas temperature ($T_k$, dotted lines) and the spin temperature ($T_s$, solid lines) as a function of redshift $z$ for accreting PBHs, 
including the accretion of baryonic matter and DM particles, with $M_{\rm PBH}=10^{2} M_{\odot}, f_{\rm PBH}=10^{-6}$ (blue lines) and $10^{-5}$ (green lines). The case with no contributions from accreting PBHs is also shown for comparison ($f_{\rm PBH}=0$, red lines). The temperature of the CMB ($T_r$, dashed line) is also shown.}
\label{fig:tem}
\end{figure}

In Fig.~\ref{fig:tem}, the changes of $T_k$ with redshift are shown for $f_{\rm PBH}=10^{-6}$ and $10^{-5}$ with $M_{\rm PBH}=10^{2}~M_{\odot}$ (dotted lines). We also plot the default case with no contributions from accreting PBHs for comparison ($f_{\rm PBH}=0$). As shown in Fig.~\ref{fig:tem}, compared with the default case, the gas temperature of the IGM increases significantly in redshift $z\lesssim 50$ due to the effect of accreting PBHs, especially for a large mass fraction of PBHs. One should  notice that in this paper we have not included the impacts of x rays, which is proportional to the star formation rate~\cite{binyue,Furlanetto:2006tf,Oh:2000zx}.

\subsection{The global 21-cm signal in the cosmic dawn including accreting PBHs and constraints on the abundance of PBHs}

In this section we review the essential parts related to the global 21-cm signal, and one can refer to, e.g., Refs.~\cite{Pritchard:2011xb,Furlanetto:2006jb} for more details. One of the important quantities related to the global 21-cm signal is the 
spin temperature, $T_s$, defined as follows~\cite{Pritchard:2011xb,Furlanetto:2006jb}:  

\beqa
\frac{n_1}{n_0}=3~\mathrm{exp}\left(-\frac{T_{\star}}{T_s}\right),
\eeqa
where $n_0$ and $n_1$ are the number densities of hydrogen atoms in triplet and singlet states, respectively. The temperature $T_{\star} = 0.068~\rm K$ corresponds to the energy changes between triplet and singlet states. The spin temperature is mainly effected by background photons, collisions between the particles, and resonant scattering of $\rm Ly\alpha$ photons (Wouthuysen-Field effect)~\cite{Pritchard:2011xb,Furlanetto:2006jb}. Taking these factors into account and with the CMB as the main background, the spin temperature can be expressed as follows~\cite{binyue,Cumberbatch:2008rh}

\beqa
T_{s} = \frac{T_{\rm CMB}+(y_{\alpha}+y_{c})T_{k}}{1+y_{\alpha}+y_{c}},
\eeqa
where $y_{\alpha}$ corresponds to the Wouthuysen-Field effect and we adopt the formula used in, e.g., Refs.~\cite{binyue,mnras,Kuhlen:2005cm}: 

\beqa
y_{\alpha} = \frac{P_{10}}{A_{10}}\frac{T_{\star}}{T_{k}}{\rm exp}\left[\frac{-0.3(1+z)^{\frac{1}{2}}}{T_{k}^{\frac{2}{3}}\left(1+\frac{0.4}{T_{k}}\right)}\right],
\eeqa
where $A_{10}=2.85\times 10^{-15}s^{-1}$ is the Einstein coefficient of hyperfine spontaneous transition. 
$P_{10}$ is the radiative de-excitation rate due to Ly$\alpha$ photons
~\cite{Pritchard:2011xb,Furlanetto:2006jb}. The factor $y_c$ involves collisions between hydrogen atoms and other particles~\cite{binyue,prd-edges,Kuhlen:2005cm,Liszt:2001kh,epjplus-2}, 

\beqa
y_{c} = \frac{(C_{\rm HH}+C_{\rm eH}+C_{\rm pH}){T_{\star}}}{A_{10}T_{k}},
\eeqa  
where $C_{\rm HH, eH, pH}$ are the deexcitation rates due to collisions and the fitted formulas can be found in Refs.~\cite{prd-edges,epjplus-2,Kuhlen:2005cm,Liszt:2001kh}. 

The differential brightness temperature, $\delta T_{21}$, relative to the CMB background, can be written as follows~\cite{Cumberbatch:2008rh,Ciardi:2003hg,prd-edges} 

\beqa
\delta T_{21} =~&&26(1-x_e)\left(\frac{\Omega_{b}h}{0.02}\right)\left[\frac{1+z}{10}\frac{0.3}{\Omega_{m}}\right]^{\frac{1}{2}} \nonumber \\
&&\times \left(1-\frac{T_{\rm CMB}}{T_s}\right)\rm mK,
\eeqa
where $\Omega_{b}$ and $\Omega_{m}$ are the density parameters of baryonic matter and DM, respectively. $h$ is the reduced Hubble constant. 

\begin{figure}
\includegraphics[width=0.5\textwidth]{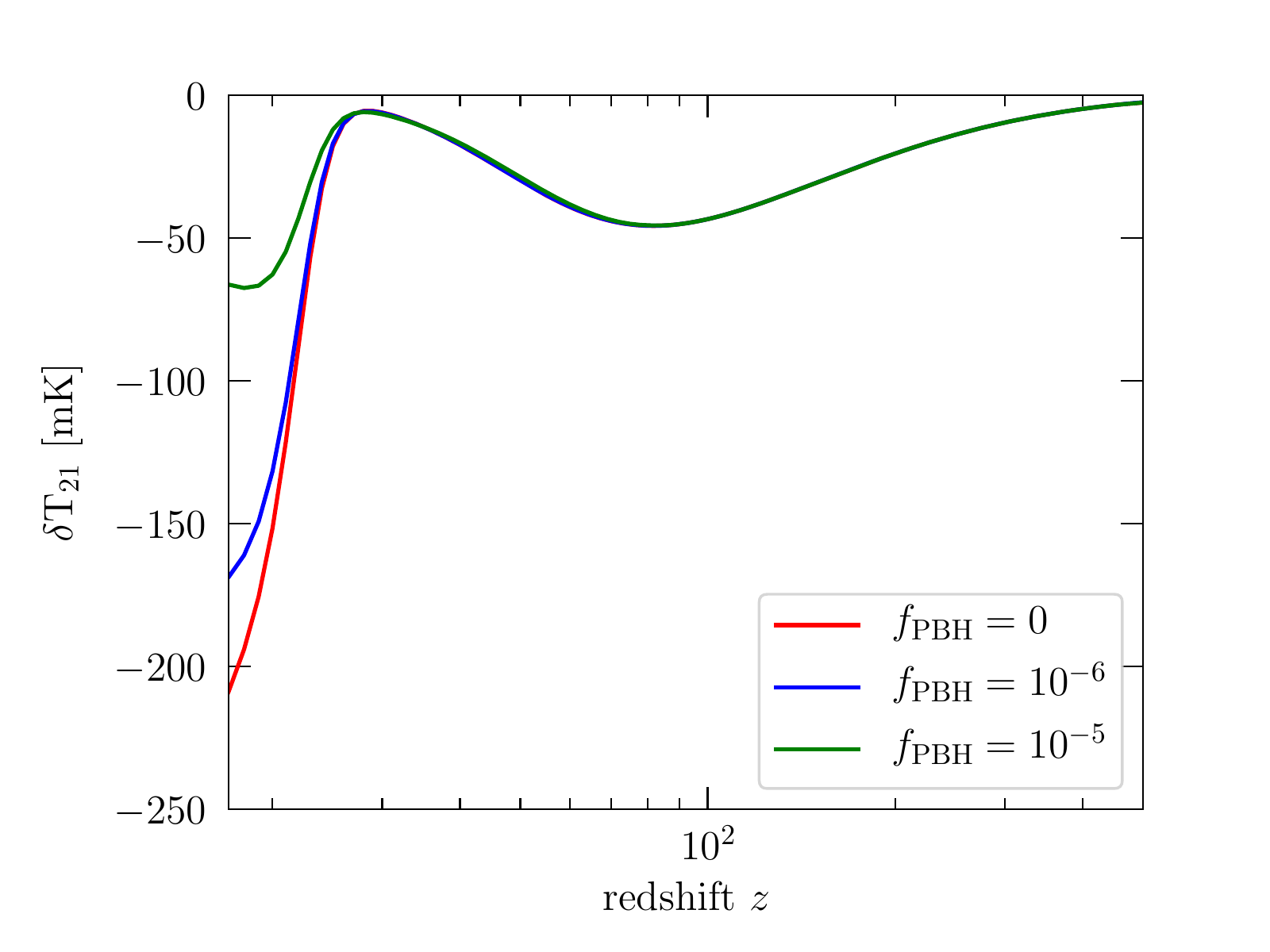}
\caption{The evolution of the differential brightness temperature $\delta T_{21}$ with redshift $z$ for accreting PBHs. 
The line style is the same as in Fig.~\ref{fig:tem}.}
\label{fig:dtb}
\end{figure}

\begin{figure}
\includegraphics[width=0.5\textwidth]{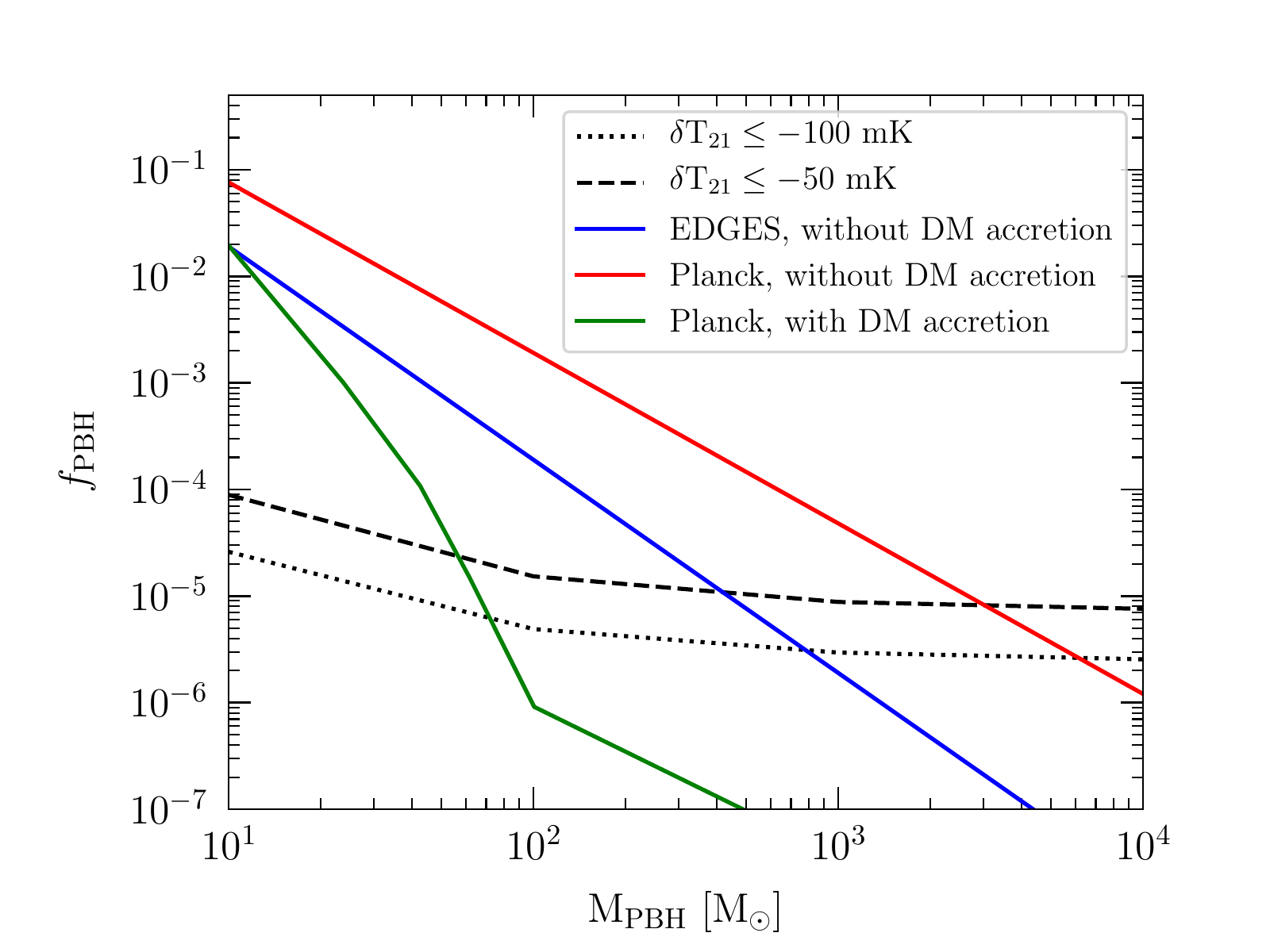}
\caption{Upper limits on the abundance of PBHs for $\delta T_{21} \lesssim -100~{\rm mK}$ (dotted line) and $-50~\rm mK$ (dashed line). For comparison, the constraints using the Planck data with (green line)~\cite{Poulin:2017bwe} and without DM particle accretion (red line)~\cite{PhysRevResearch.2.023204}, and the global 21-cm signal without DM particle accretion (blue line)~\cite{Hektor:2018qqw} are also shown. For the Planck limits without DM particle accretion, we have used the fitted formula in Ref.~\cite{Poulin:2017bwe}, $f_{\rm PBH}<\left(2M_{\odot}/M\right)^{2.6}\left(0.01/\lambda\right)^{1.6}$ with $\lambda = 0.01$. For the EDGES limits, we have used the fitted formular in Ref.~\cite{Hektor:2018qqw}, $f_{\rm PBH} < C(\beta)\left(0.15/f_{E}\right)\left(\lambda/0.01~M_{\rm PBH}/10M_{\odot}\right)^{-1-\beta}$ with $\lambda =0.01$, 
$f_{E}=0.15$, $\beta=1$ and $C(\beta) = 0.019\beta^{2.5}$.}
\label{fig:fraction}
\end{figure}

The changes of the spin temperature $T_s$ with redshift are shown in Fig.~\ref{fig:tem} for $f_{\rm PBH}=10^{-6}$ and $10^{-5}$ with $M_{\rm PBH}=10^{2}~M_{\odot}$ (solid lines). For the considered models of accreting PBHs here, the significant deviations of $T_s$ from the default case appear in redshit $z\lesssim 40$. The changes of the differential brightness temperature $\delta T_{21}$ with redshift are shown in Fig.~\ref{fig:dtb}. For our considered models of accreting PBHs, the amplitude of the global 21-cm signal in the cosmic dawn decreases compared with the default case ($f_{\rm PBH}=0$). By requiring the differential brightness temperature $\delta T_{21} \lesssim -100$ and $-50~\rm mK$, the upper limits on the mass fraction of PBHs are obtained and shown in Fig.~\ref{fig:fraction}. For $M_{\rm PBH} =10^{4}~M_{\odot}$, the limits on the fraction of PBHs are $f_{\rm PBH}\lesssim 2.6\times 10^{-6}$ ($7.6\times 10^{-6}$) for $\delta T_{21} \lesssim -100$ ($-50$) $\rm mK$. For lower mass, e.g, $M_{\rm PBH} = 10~M_{\odot}$, the limits are $f_{\rm PBH}\lesssim 2.6\times 10^{-5}$ ($8.9\times 10^{-5}$) for $\delta T_{21} \lesssim -100$ ($-50$) $\rm mK$. For comparison, the upper limits obtained using the Planck data with~\cite{Poulin:2017bwe} and without DM accretion~\cite{PhysRevResearch.2.023204} and the global 21-cm signal detected by EDGES without DM accretion~\cite{Hektor:2018qqw} are also shown. One should notice that in Refs.~\cite{Poulin:2017bwe,PhysRevResearch.2.023204,Hektor:2018qqw} a constant value of the accretion parameter $\lambda$ has been used for getting the upper limits on $f_{\rm PBH}$, while a different one, Eq.~(\ref{eq:lambda}), has been used for our calculations. Therefore, as shown in Fig.~\ref{fig:fraction}, different trends in $f_{\rm PBH}$ are partly due to using the different forms of $\lambda$. Another feature shown in Fig.~\ref{fig:fraction} is that the constraints from Ref.~\cite{Hektor:2018qqw} are stronger than our limits for $M_{\rm PBH} \gtrsim 10^{3}M_{\odot}$. The main reason is the use of the different accretion models. For example, in Ref.~\cite{Hektor:2018qqw}, the accretion parameter $\lambda$ is treated as a 
free parameter ($0.01\lesssim \lambda \lesssim 0.001$) and only one of the limit lines is shown in Fig.~\ref{fig:fraction} 
(blue line, $\lambda =0.01$). In fact, as shown 
in Fig.~5 of Ref.~\cite{Hektor:2018qqw}, the limits are weakened (strengthened) for a smaller (larger) 
value of $\lambda$ for fixed $\beta$. For $\lambda=0.001$ ($\beta=1$), the upper limit on 
$f_{\rm PBH}$ is $\sim 2\times 10^{-4}$ for $M_{\rm PBH} = 10^{3}M_{\odot}$, which is about an order of magnitude weaker 
than our limits ($\delta T_{\rm 21} \lesssim 50~\rm mK$). 
On the other hand, in order to get the conservative upper limits on $f_{\rm PBH}$, 
the value of the radiative efficiency used here is smaller than that used in Ref.~\cite{Hektor:2018qqw}, and a larger 
one will strengthen our limits.

In addition to the changes of the accretion rate, the mass distribution of PBHs can also be changed due to DM particle accretion. Generally, as shown in Ref.~\cite{DeLuca:2020fpg}, the changes of the mass distribution of PBHs result in weakening the upper limits on the mass fraction of PBHs depending on redshift $z_{\rm cutoff}$, after which the mass accretion can be ignored due to the formation of 
the large scale structures.

\section{conclusions}
We have investigated the influence of accreting PBHs on the global 21-cm signal in the cosmic dawn, including the accretion of baryonic matter and DM particles. The accretion rates of PBHs are different in the cases with and without DM particle accretion. Based on the recent detection of the global 21-cm signal by EDGES, by requiring the differential brightness temperature $\delta T_{21}~\lesssim -100$ ($-50$)~$\rm mK$, we obtained the upper limits on the abundance of PBHs for the mass range $10\lesssim M_{\rm PBH} \lesssim 10^{4} M_{\odot}$. For the considered models of accreting PBHs here, the limits are $f_{\rm PBH}\lesssim 2.6\times10^{-5}$ ($2.6\times10^{-6}$) for $M_{\rm PBH}=10$ ($10^{4}$)~$M_{\odot}$ for $\delta T_{21}\lesssim -100~\rm mK$. The limits are weakened by a factor of $\sim 3$ for $\delta T_{21}\lesssim -50~\rm mK$. 
Compared with previous works, where the limits are obtained using the Planck data (with and without DM accretion) 
and the global-21 cm signal (without DM accretion), the limits are stronger for the mass range $10\lesssim M_{\rm PBH}\lesssim 50~M_{\odot}$.

\section{Acknowledgements}
Y. Yang thanks Dr. Bin Yue and Xiaoyuan Huang for the helpful discussions. This work is supported in part by the 
National Natural Science Foundation of China (under Grant No.11505005). 
Y. Yang is supported by the Youth Innovations and Talents Project of Shandong Provincial Colleges and Universities (Grant No. 201909118).
\

\bibliographystyle{apsrev4-1}
%

\end{document}